\documentclass{emulateapj}
\usepackage{apjfonts}

\newcommand{\mc}{\multicolumn}

\newcommand{\ltsimeq}{\raisebox{-0.6ex}{$\,\stackrel 
        {\raisebox{-.2ex}{$\textstyle <$}}{\sim}\,$}} 
\newcommand{\gtsimeq}{\raisebox{-0.6ex}{$\,\stackrel 
        {\raisebox{-.2ex}{$\textstyle >$}}{\sim}\,$}} 
\newcommand{\umu}{\mu}

\newcommand{\civ}{C\,{\sc iv}}
\newcommand{\cii}{[C\,{\sc ii}]}
\newcommand{\ciii}{C\,{\sc iii}]}
\newcommand{\mgi}{Mg\,{\sc i}}
\newcommand{\mgii}{Mg\,{\sc ii}}
\newcommand{\feii}{Fe\,{\sc ii}}
\newcommand{\lya}{Ly\,$\alpha$}
\newcommand{\lyb}{Ly\,$\beta$}
\newcommand{\lyg}{Ly\,$\gamma$}
\newcommand{\nv}{N\,{\sc v}}

\newcommand{\oiii}{[O\,{\sc iii}]}

\newcommand{\oi}{O\,{\sc i}}

\newcommand{\hii}{H\,{\sc ii}}
\newcommand{\fhi}{$f_{\rm HI}$}

\def\co21{CO\,(2-1)}

\shorttitle{Four quasars above redshift 6 discovered by the CFHQS}
\shortauthors{Willott et al.}

\begin{document}


\title{Four quasars above redshift 6 discovered by the Canada-France High-$z$ Quasar Survey}


\author{
Chris J. Willott\altaffilmark{1},
Philippe Delorme\altaffilmark{2},
Alain Omont\altaffilmark{3},
Jacqueline Bergeron\altaffilmark{3},
Xavier Delfosse\altaffilmark{2},
Thierry Forveille\altaffilmark{2},
Loic Albert\altaffilmark{4},
C\'eline Reyl\'e\altaffilmark{5},
Gary J. Hill\altaffilmark{6},
Michael Gully-Santiago\altaffilmark{6},
Phillip Vinten\altaffilmark{1},
David Crampton\altaffilmark{7},
John B. Hutchings\altaffilmark{7},
David Schade\altaffilmark{7},
Luc Simard\altaffilmark{7},
Marcin Sawicki\altaffilmark{8},
Alexandre Beelen\altaffilmark{9},
and Pierre Cox\altaffilmark{10}
}

\altaffiltext{1}{University of Ottawa, Physics Department, 150 Louis Pasteur, MacDonald Hall, Ottawa, ON K1N 6N5,  Canada; cwillott@uottawa.ca}
\altaffiltext{2}{Laboratoire d'Astrophysique, Observatoire de Grenoble, Universit\'e J. Fourier, BP 53, F-38041 Grenoble, Cedex 9, France}
\altaffiltext{3}{Institut d'Astrophysique de Paris-CNRS, 98bis Boulevard Arago, F-75014, Paris, France}
\altaffiltext{4}{Canada-France-Hawaii Telescope Corporation, 65-1238 Mamalahoa 
Highway, Kamuela, HI96743, USA}
\altaffiltext{5}{Institut Utinam, Observatoire de Besan\c{c}on, BP1615, 25010 Besan\c{c}on Cedex, France}
\altaffiltext{6}{McDonald Observatory, University of Texas at Austin, 1 University Station, Austin, TX 78712, USA}
\altaffiltext{7}{Herzberg Institute of Astrophysics, National Research
Council, 5071 West Saanich Rd, Victoria, BC V9E 2E7, Canada}
\altaffiltext{8}{Department of Astronomy and Physics, St. Mary's  University, 923 Robie St, Halifax, NS  B3H 3C3, Canada}
\altaffiltext{9}{Institut d'Astrophysique Spatiale, Universit\'e Paris-Sud, 91405 Orsay Cedex, France} 
\altaffiltext{10}{Institute de Radioastronomie Millimetrique, St. Martin d'Heres, F-38406, France}

\begin{abstract}

The Canada-France High-$z$ Quasar Survey (CFHQS) is an optical survey
designed to locate quasars during the epoch of reionization. In this
paper we present the discovery of the first four CFHQS quasars at
redshift greater than 6, including the most distant known quasar,
CFHQS\,J2329-0301 at $z=6.43$.  We describe the observational method
used to identify the quasars and present optical, infrared, and
millimeter photometry and optical and near-infrared spectroscopy.  We
investigate the dust properties of these quasars finding an unusual
dust extinction curve for one quasar and a high far-infrared
luminosity due to dust emission for another. The mean millimeter
continuum flux for CFHQS quasars is substantially lower than that for
SDSS quasars at the same redshift, likely due to a correlation with
quasar UV luminosity. For two quasars with sufficiently high
signal-to-noise optical spectra, we use the spectra to investigate the
ionization state of hydrogen at $z>5$. For CFHQS\,J1509-1749 at
$z=6.12$, we find significant evolution (beyond a simple extrapolation
of lower redshift data) in the Gunn-Peterson optical depth at
$z>5.4$. The line-of-sight to this quasar has one of the highest known
optical depths at $z\approx 5.8$. An analysis of the sizes of the
highly-ionized near-zones in the spectra of two quasars at $z=6.12$
and $z=6.43$ suggest the IGM surrounding these quasars was
substantially ionized before these quasars turned on. Together, these
observations point towards an extended reionization process, but we
caution that cosmic variance is still a major limitation in $z>6$
quasar observations.

\end{abstract}


\section{Introduction}

The earliest observed epoch of the universe is 380\,000 years after
the Big Bang when protons and electrons combined into neutral
hydrogen. The radiation from this epoch, observed as the Cosmic
Microwave Background (CMB), shows that the universe was homogeneous
with fractional density fluctuations of only $\sim 10^{-4}$. For the
following few hundred million years there were no stars or
galaxies. This period is often referred to as the Dark Ages because of
the absence of sources of light (Rees 1999).

Eventually these small density fluctuations in the dark matter
distribution grew into dark matter halos where gas could cool and form
galaxies. As stars and black holes formed within galaxies they emitted
copious amounts of ultraviolet (UV) radiation. Many of these photons
escaped their host galaxies and ionized hydrogen in the intergalactic
medium (IGM). When sufficient sources of ionizing photons had formed,
they provided an ionizing background radiation field high enough to
ionize all the diffuse neutral hydrogen in the IGM. The details of
reionization are sensitive to the luminosity function of ionizing
sources, the fraction of ionizing photons that escape from them, the
ionizing source spectra and the clumpiness of the IGM
(Miralda-Escud\'e et al. 2000).

Quasars are useful probes of reionization because they can be detected
at very high redshift and have strong intrinsic UV radiation that is
absorbed at the resonant Lyman lines by neutral hydrogen. The presence
of significant flux in the \lya\ forest of quasars at $z<6$ indicates
that the IGM is almost completely reionized by this epoch (Fan et
al. 2000; Fan et al. 2001; Songaila 2004). The discovery of
Gunn-Petersen troughs in the spectra of quasars at $6<z<6.42$ have
been interpreted as evidence that we are approaching the epoch of
reionization (Becker et al. 2001; White et al. 2003; Fan et
al. 2006a). Fan et al. (2006b) analyzed a sample of 19 SDSS quasars at
$z>5.7$ and found that the mean Gunn-Peterson optical depth and its
variance increase sharply away from the low redshift extrapolation at
$z>5.7$, consistent with the overlap phase of reionization occurring
at $z \approx 6.1$ (Gnedin \& Fan 2006).  However, it has also been
argued that these data are compatible with much earlier reionization
(Songaila 2004; Becker et al. 2007).

The sizes of highly-ionized near-zones around quasars are sensitive to
the neutral fraction of hydrogen before the quasar turned on (Cen \&
Haiman 2000). However, the interpretation of these results is complex
and various authors have come to a wide range of conclusions on the
basis of the same SDSS quasar spectra (Wyithe \& Loeb 2004; Wyithe,
et al. 2005; Yu \& Lu 2005; Fan et al. 2006b; Bolton \&
Haehnelt 2007a,b; Mesinger \& Haiman 2007; Alvarez \& Abel 2007). Despite
these uncertainties, this could be a powerful probe at high neutral
fractions where traditional Gunn-Peterson measurements are
insensitive. Becker et al. (2006) studied \oi\ absorption up to
$z=6.42$ and did not find an \oi\ forest suggesting that these lines
of sight are either largely unenriched or significantly ionized.

Simulations predict that reionization is an inhomogeneous process due
to the existence of discrete, clustered ionizing sources and an
inhomogeneous IGM (e.g. Barkana \& Loeb 2001).  Cosmic variance in
transmission due to large-scale density inhomogeneities is extremely
large on proper scales up to 20\,Mpc at $z=5.5$ and rapidly increases
at earlier times before full reionization (Wyithe \& Loeb 2006). The
SDSS presently contains only a handful of quasars at a high enough
redshift ($z>6.1$) to probe a large path length of the $z > 6$
IGM. From Fan et al. (2006a) and refs therein, the path length at $z >
6$ presently probed by the SDSS quasars is only $\sim 40$\, proper Mpc
(excluding the broad absorption line quasar SDSS\,J1048+4607, whose
broad absorption lines may be confused with IGM absorption).
Observations of SDSS quasars at $z>6.2$ show evidence for significant
cosmic variance with a substantial range of foreground IGM absorption
(Oh \& Furlanetto 2005). More quasars are required to overcome cosmic
variance at $z>6$.

Observations of high-redshift quasars are important not only for studying
reionization, but also for the formation of supermassive black holes and
their host galaxies.
The high luminosities and broad line widths of the most distant
quasars require black holes with masses greater than $10^9 M_\odot$
(Fan et al. 2001; Willott et al. 2003). Forming such
massive black holes within the first billion years of the universe
provides a challenge to models of galaxy formation, black hole
formation and black hole growth (Bromley et al. 2004;
Yoo \& Miralda-Escud\'e 2004; Shapiro 2005; Volonteri \& Rees 2005;
King \& Pringle 2006).

It is well established that there is a tight relation between the mass
of the black hole and the spheroid in local galaxies (Magorrian et
al. 1998). However the evolution of this relation with redshift is not
well known and attempts are now being made to determine the relation
up to $z=3$ because it is a crucial component of galaxy evolution
models (Shields et al. 2003; Adelberger \& Steidel 2005; Alexander et
al. 2005; Peng et al. 2006; McLure et al. 2006). It is also important
to push this relation out to $z \sim 6$ where the earliest known black
holes exist. Studies of millimeter radiation show that at least 40\% of $z
\sim 6$ quasars have high far-infrared luminosities indicative of high
star formation rates of $\rm \sim 1000 \, M_\odot \, yr^{-1}$ (Wang et
al. 2007). SDSS\,J1148+5251 has resolved CO emission making it the
only $z>6$ quasar with a dynamical mass measurement (Walter et
al. 2004). Curiously, the mass inferred is an order of magnitude less
than expected if there is no evolution in the black hole -- spheroid
relation.

To date, only 10 quasars have been discovered at redshift greater than
or equal to 6 (Fan et al. 2006a and refs therein; McGreer et al. 2006).
To discover more very distant quasars which can be used as probes of
reionization and the early growth of massive black holes and galaxies,
we are carrying out the Canada-France High-z Quasar Survey
(CFHQS). This project uses the 1 square degree imager MegaCam at the
Canada-France-Hawaii Telescope. The survey plans to cover $\sim 900$
square degrees in 4 optical filters $g',r',i',z'$ to a limiting depth
of $z'\approx 22.5$. This is more than 2 magnitudes fainter than the
SDSS survey leading to the discovery of correspondingly lower
luminosity quasars. Lower luminosity quasars are expected to have
lower black hole masses which could be due to the black holes being
younger than those in SDSS quasars. The survey is carried out in areas
of the sky which already have MegaCam imaging in some filters as part
of the CFHT Legacy Survey\footnotemark\
\footnotetext{http://www.cfht.hawaii.edu/Science/CFHTLS} or RCS2
Survey\footnotemark.\footnotetext{http://www.rcs2.org} As discussed
below, the survey is also extremely effective at finding ultracool
brown dwarfs. In this paper, we present the first high-redshift
quasars to be discovered in the CFHQS.

In the following section we describe how the quasars were discovered
from imaging and spectroscopic observations. Sec.\,3 presents the
properties of the four quasars. Sec.\,4 deals with observations at other
wavelengths including  millimeter observations. In Sec.\,5 we
discuss constraints on the ionization state of hydrogen at high
redshift from these quasars. Sec.\,6 presents our conclusions.

All optical and near-IR magnitudes in this paper are on the AB
system. Cosmological parameters of $H_0=70~ {\rm km~s^{-1}~Mpc^{-1}}$,
$\Omega_{\mathrm M}=0.27$ and $\Omega_\Lambda=0.73$ (Spergel et
al. 2007) are assumed throughout. The convention for spectral indices,
$\alpha_\nu$, is that $f_\nu \propto \nu^{-\alpha_\nu}$.


\section{Observations}

\subsection{Imaging}

The four quasars whose discovery we report here were found in the
first two years of optical imaging of the CFHQS.  The total area
covered at this time is $\approx 400$ square degrees to typical limits
of $z'\approx 22.5, i'\approx 24.5$.  However, because follow-up
observations are not complete, we expect this region to contain more
quasars than discovered so far and no conclusions on the space density
of low luminosity quasars and shape of the luminosity function at
$z=6$ are available at this time.

We only briefly describe the processing and photometry of the imaging
observations in this paper. A much fuller description is given in
Delorme et al. (2007). Pre-processing of the MegaCam images is carried
out at CFHT using the ELIXIR pipeline.  This removes the instrumental
effects from the images. We then run our own algorithms to improve the
astrometry and check the photometry. Finally, we stack the images (for
those cases where there is more than one exposure at a given position)
and register the images in the four different filters.  Photometry is
carried out using our own adaption of Sextractor (Bertin \& Arnouts
1996) which uses dual image multiple psf-fitting (Delorme \& Bertin in
prep.) to optimize the signal-to-noise ratio (S/N) of point sources.
 
Candidate quasars are initially identified on the optical images as
objects which have very high $i'-z'$ colors. In Willott et al. (2005a)
we described in detail our method for identifying high-$z$ quasars and
brown dwarfs using MegaCam optical plus near-IR imaging. As shown in
that paper, photometric noise cause many M stars (with intrinsic
colors of $0.5 < i'-z' < 1$) to be scattered into the region of the
diagram at $i'-z' > 1.5$ where we would expect to find only L or T
dwarfs and quasars. The huge number of these M stars would require a
lot of telescope time for complete follow-up. Therefore we now limit
our survey to objects observed to be redder than $i'-z' > 1.7$, with
stronger priority for follow-up given to redder objects. In Willott et
al. (2005a), we showed that changing the criterion from $i'-z'>1.5$ to
$i'-z'>1.7$ only loses us quasars at $z<5.9$, but reduces the number
of M dwarf contaminants by $>80$\%. Similarly, imposing a cut at
$i'-z'>1.8$ or even $i'-z'>2$ would have very little effect on the
number of $z>6$ quasars discovered. All of the quasars presented in
this paper were observed to have $i'-z'>2$.

\begin{table*}
\begin{center}
\caption{\label{tab:photom} Positions and photometry for the CFHQS quasars.} 
\begin{tabular}{clccccc}
\hline
\hline
\mc{1}{c}{Quasar} &\mc{1}{c}{RA and DEC (J2000.0)} &\mc{1}{c}{$i'$ mag} &\mc{1}{c}{$z'$ mag}&\mc{1}{c}{$J$ mag} &\mc{1}{c}{$i'-z'$} &\mc{1}{c}{$z'-J$} \\
\hline
CFHQS\,J003311-012524 &  00:33:11.40  -01:25:24.9 &  $24.82 \pm 0.15$ & $22.41 \pm 0.08$  & $21.58 \pm 0.20$ & $2.41 \pm 0.17$ & $0.74 \pm 0.22$ \\
CFHQS\,J150941-174926 &  15:09:41.78  -17:49:26.8 &  $23.11 \pm 0.05$ & $20.26 \pm 0.02$  & $19.68 \pm 0.10$ & $2.86 \pm 0.05$ & $0.58 \pm 0.10$ \\
CFHQS\,J164121+375520 &  16:41:21.64  +37:55:20.5 &  $23.69 \pm 0.20$ & $21.31 \pm 0.04$  & $21.24 \pm 0.14$ & $2.38 \pm 0.20$ &  $0.07 \pm 0.15$ \\ 
CFHQS\,J232908-030158 &  23:29:08.28  -03:01:58.8 &   $>25.08$\tablenotemark{a}        & $21.76 \pm 0.05$  & $21.56 \pm 0.25$ & $>3.32$         &  $0.18 \pm 0.25$ \\
\hline
\end{tabular}
\tablenotetext{a}{Where not detected at $> 2\sigma$ significance, a $2\sigma$ lower limit is given.}
\end{center}
\end{table*}

To separate quasars from brown dwarfs we employ near-infrared $J$-band
imaging (e.g. Fan et al. 2001). Quasars at $z<6.5$ appear relatively
blue in $z'-J$, whereas L and T dwarfs are very red. Near-IR imaging
was carried out at several telescopes and the observations, processing
and photometry are described in Delorme et al. (2007).

\begin{figure}
\hspace{-0.1cm}
\resizebox{0.48\textwidth}{!}{\includegraphics{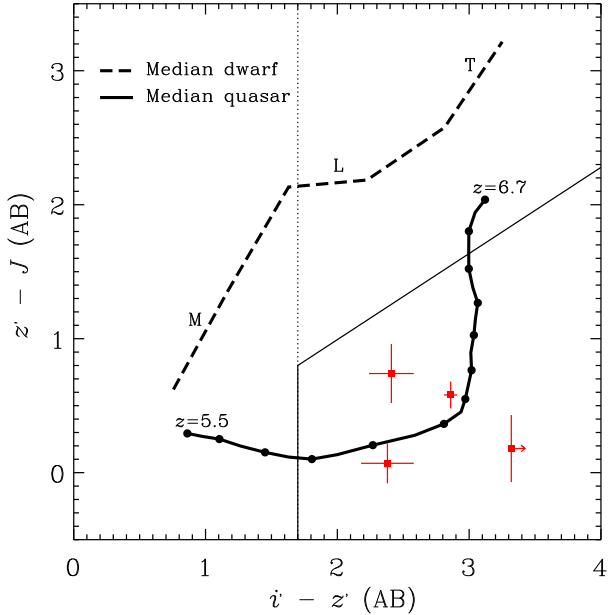}}
\caption{Color-color diagram showing the selection region for high-$z$
  quasars (to the lower-right of the thin solid line). The thin dotted line shows the optical criterion of $i'-z'>1.7$ which also finds many L and T dwarfs. The thick solid line is the median cloned quasar track for quasars at $5.5<z<6.7$ with a mark every $\delta z=0.1$ (from Willott et al. 2005a). The thick dashed line is the median brown dwarf track for types M, L and T. The four squares with error bars are the four new $z>6$ quasars. Colors of the brown dwarfs discovered by this survey will be presented in Delorme et al. (2007).  
\label{fig:izj}
}
\end{figure}

\begin{figure}
\hspace{-0.2cm}
\resizebox{0.48\textwidth}{!}{\includegraphics{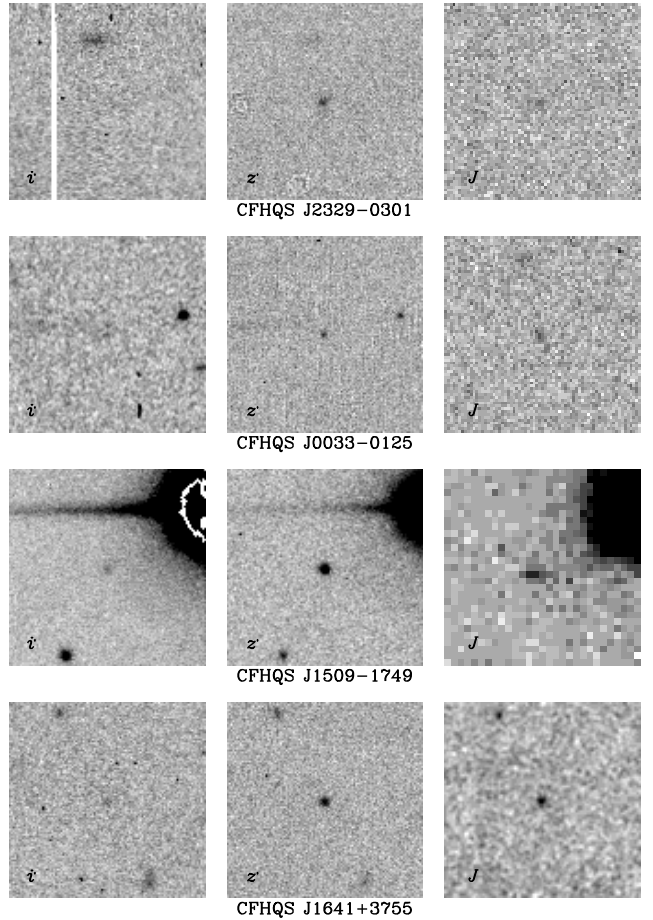}}
\caption{Images in the $i'$, $z'$ and $J$ filters centred on the four
  CFHQS quasars. Each image covers $20'' \times 20''$. The images are
  oriented with north up and east to the left.
\label{fig:cutouts}
}
\end{figure}

Fig.\,\ref{fig:izj} shows the $i'-z'$ vs $z'-J$ color-color diagram
for the newly discovered quasars. Also shown are lines corresponding
to the expected median colors of quasars at redshifts $5.5<z<6.7$ and
the median track for brown dwarfs discovered by 2MASS and SDSS. The
expected quasar color track is based on ``cloning'' of 180 SDSS
quasars at $z=3$ (see Willott et al. 2005a for full details). Three of the quasars lie
very close to the median quasar track. The exception is
CFHQS\,J2329-0301 which is much redder in $i'-z'$ and bluer in $z'-J$
than expected given its redshift. The reason for this will be
discussed in Sec.\,\ref{indiv}. Positions and photometry for the four quasars are given in Table\,\ref{tab:photom}.

$20'' \times 20''$ images centred on the quasars are shown in
Fig.\,\ref{fig:cutouts}. All quasars are clearly detected at
$z'$-band, but most have only faint detections at $i'$ and
$J$. CFHQS\,J2329-0301 is the only quasar not detected at
$i'$. CFHQS\,J1509-1749 is fortuitously located only $12''$ from a
very bright ($R=12.6$) star and is ideally suited to natural guide
star adaptive optics observations. Finding charts in the $z'$-band
over a wider field are presented in the appendix.

\begin{table*}
\begin{center}
{\caption{\label{tab:specobs} Optical spectroscopy observations of CFHQS quasars.}}
\begin{tabular}{ccclccccc}
\hline
\hline
\mc{1}{c}{Quasar} &\mc{1}{c}{Redshift} &\mc{1}{c}{Telescope} &\mc{1}{c}{Date~~~~~~~}&\mc{1}{c}{Resolving} &\mc{1}{c}{Slit Width} &\mc{1}{c}{Exp. Time} &\mc{1}{c}{Seeing}  &\mc{1}{c}{$M_{1450}$} \\
\mc{1}{c}{ }      &\mc{1}{c}{$z$}      &\mc{1}{c}{ }
&\mc{1}{c}{ }   &\mc{1}{c}{Power}     &\mc{1}{c}{(Arcsec)}
&\mc{1}{c}{(s)}       &\mc{1}{c}{(Arcsec)} &\mc{1}{c}{ } \\
\tableline
CFHQS\,J0033-0125 & 6.13 & Gemini  & 2006 Nov 26  & 1300 & 1.0 & 3600
& 0.6 & $-25.03$  \\
CFHQS\,J1509-1749 & 6.12 & Gemini  & 2006 May 26  & 2100 & 0.5 & 7200 & 0.7 & $-26.98$ \\
CFHQS\,J1641+3755 & 6.04 & HET     & 2006 July 26 + Sep 19 & 400 & 2.0 & 5700 & 1.9
&  $-25.48$ \\
CFHQS\,J2329-0301 & 6.43 & Gemini  & 2006 Nov 26  & 1300 & 1.0 & 3600 & 0.7 & $-25.23$ \\
\hline
\end{tabular}
\end{center}
\end{table*}

\subsection{Spectroscopy}
\label{spec}

Optical spectroscopy of candidate quasars was carried out using the
Gemini Multi-Object Spectrograph (GMOS; Hook et al. 2004) at the
Gemini South Telescope and the Marcario Low Resolution Spectrograph
(LRS; Hill et al. 1998) at the Hobby-Ebberly Telescope. The details of
these observations are given in Table\,\ref{tab:specobs}.

The GMOS spectroscopy was carried out using the nod-and-shuffle mode
to minimize sky subtraction residuals at the red end of the spectra.
Most of the reductions of the Gemini spectra were performed using
tasks in the {\sc IRAF} Gemini package, which were specifically
designed for GMOS. The first step is removal of the bias level using a
bias frame constructed from many ($>10$) bias observations (with the
same binning) carried out during the same month. Then the sky was
removed from the spectra by subtracting a spectrum of the sky obtained
through exactly the same light path. GMOS consists of three separate
chips, so the next step was to mosaic together the data on the three
chips. Each object was observed in more than one exposure, so all
exposures of each object were registered in the spectral direction
using the location of night sky emission lines (in unsubtracted
images) and in the spatial direction using the location of the quasar
spectrum or that of a bright star located along the slit. The
registered exposures were combined using bad pixel masks (including
charge traps) and a sigma clipping algorithm. The exposures of
CFHQS\,J0033-0125 and CFHQS\,J2329-0301 each had only four spectra to
combine, so a few cosmic rays survived and were manually edited out of
the final two-dimensional spectra. Wavelength calibration was achieved
using the night sky spectrum from the unsubtracted images. The quasar
spectra were then extracted from the two-dimensional spectra.

The LRS data reduction followed similar procedures. Cosmic rays were
removed from the individual frames prior to co-addition with
L.A.Cosmic (van Dokkum 2001). The wavelength scale was set and the
distortion removed from the frames with a combination of cadmium and
neon lines. A 4th order fit to the wavelength scale yields an accuracy
of 1.0\,\AA\ or better from $6500 - 9100$\,\AA. Incomplete fringing
removal, beating with the night sky lines, causes some residual
positive features in the final spectrum. These are described further
in Sec.\,\ref{indiv}.

For the GMOS and LRS spectra, a relative flux calibration was applied
using observations of a spectrophotometric standard star. Due to slit
losses, absolute flux calibration was achieved by passing the spectra
through the CFHT MegaCam $z'$ filter + CCD response curves and
normalizing to match the observed $z'$ magnitudes.

\begin{figure*}
\resizebox{0.95\textwidth}{!}{\includegraphics{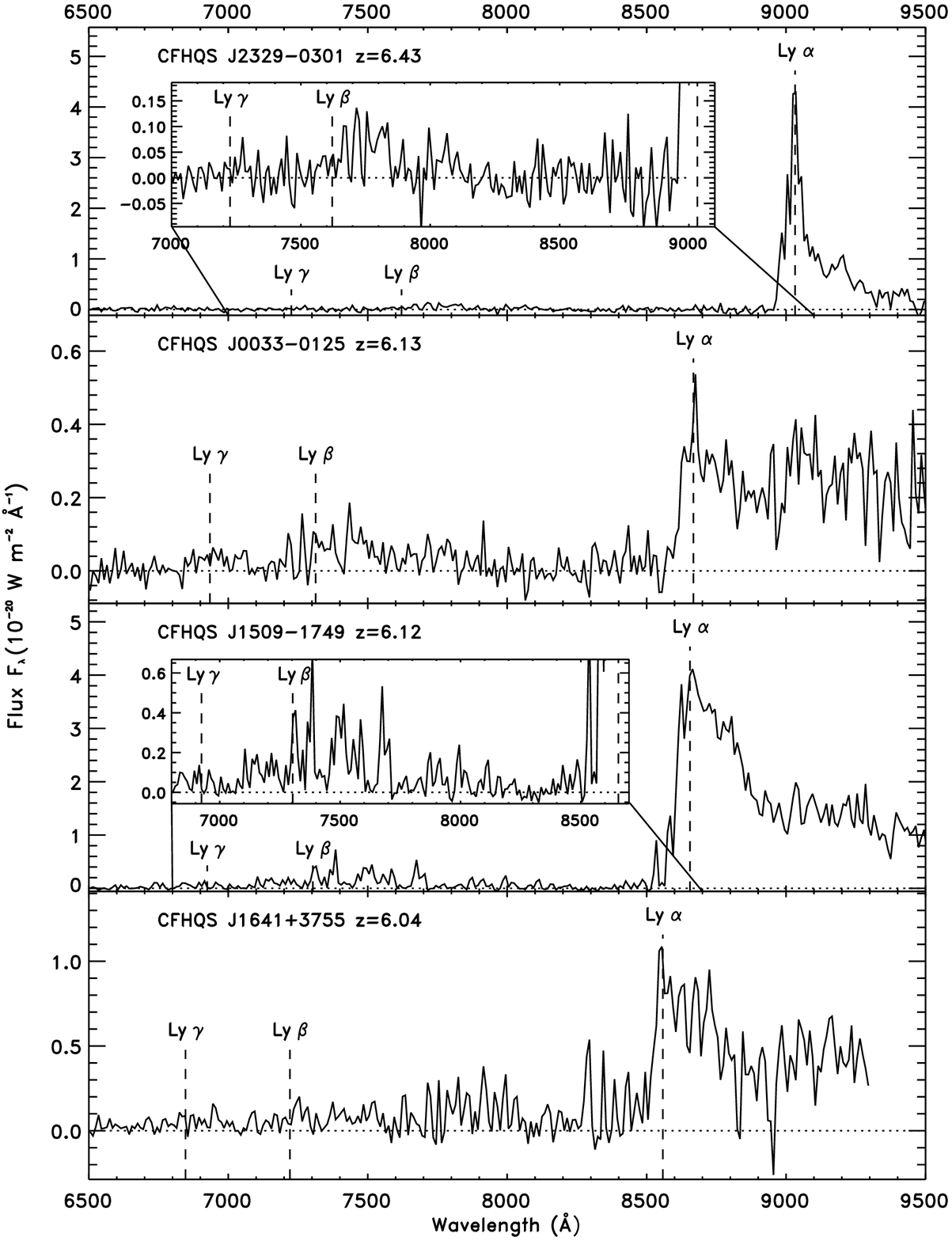}}
\caption{Optical spectra of the four newly discovered quasars. The
expected locations of \lya, \lyb\ and \lyg\ are marked with dashed
lines. The two quasars with the highest S/N spectra have inset panels
which show the \lya\ and \lyb\ forests on an expanded scale. Note the
huge variation in the strength and shape of the \lya\ lines of the
four quasars. All spectra are binned in 10\AA\ pixels. The spectrum of
CFHQS\,J1641+3755 is contaminated by night sky residuals at $\lambda >
7600$\,\AA.
\label{fig:spec}
}
\end{figure*}

The optical spectra of the four quasars are shown in
Fig.\,\ref{fig:spec}. In all four cases there is an unambiguous
spectral break close to the \lya\ emission line, indicating the object
is a $z>6$ quasar. All of the quasars except CFHQS\,J0033-0125 have 
strong \lya\ emission lines.

Note that the spectra have a wide range of S/N due to the
spread of 2 magnitudes in brightness and different telescopes and exposure
times used.  Therefore we leave a detailed discussion of the spectra
to individual sections about each quasar in Sec.\,\ref{indiv}.

Near-infrared spectroscopy of CFHQS\,J1509-1749 was performed using
the Gemini Near Infra-Red Spectrograph (GNIRS; Elias et al. 2006) at
Gemini-South on 2006 March 14. Conditions were photometric with
0.8 arcsec seeing. GNIRS was used in the cross-dispersed mode with
the 32 lines mm$^{-1}$ grating and the short camera. With a 0.675
arcsec wide slit, this setup gives the full near-IR spectrum from $0.8$ to
$2.4\,\umu$m at a constant resolution of $R=900$. The total integration
time of 3300\,s was split into 11 equal exposures to enable background
subtraction and cosmic ray rejection.

The GNIRS spectrum was reduced with the {\sc IRAF} Gemini package. The
individual frames were flat-fielded and then background subtracted
using neighbouring frames with the target shifted along the slit by 3
arcsec. The spectra were rectified and then individual orders were
extracted from the frames. The 11 individual frames for each order
were then registered and combined using a bad pixel mask and cosmic
ray rejection. Each order was wavelength calibrated with the argon arc
lamps. For orders 3 to 6, the wavelength calibration was refined using
sky emission lines. The spectra were flux calibrated using a spectrum
of an A0 type standard star. One-dimensional spectra were extracted
from each order and combined. Absolute flux calibration was achieved
by passing the spectrum through the $J$ filter response curve and
normalizing to match the observed $J$ magnitude. The near-IR spectrum
of CFHQS\,J1509-1749 is shown in Fig.\,\ref{fig:gnirs} and discussed in
Sec.\,\ref{indiv}.

\section{Notes on individual quasars}
\label{indiv}

\subsection{CFHQS\,J2329-0301}

The spectrum of this quasar shows a very high equivalent width \lya\
emission line. The line has a strong narrow peak at $\lambda=9029 \pm
2$\,\AA. The FWHM of this narrow peak is only $600 \,{\rm
km\,s}^{-1}$. Taking this narrow peak to be the systemic redshift of
the quasar (for reasons discussed below) would give a redshift of
$z=6.427 \pm 0.002$. This makes CFHQS\,J2329-0301 the most distant
known quasar, surpassing SDSS\,J1148+5251 at $z=6.419 \pm 0.001$
(Bertoldi et al. 2003b; Walter et al. 2003). The emission line has a
broad base with a precipitous drop-off on the blue side at a \lya\
redshift of $z=6.37$. The spectral extent of \lya\ emission and
transmitted flux in the Lyman forest will be discussed further in
Sec.\,\ref{igm}.

We fit a power-law continuum curve by eye to the optical spectrum
redward of \lya\ ($>9300$\,\AA) by fixing the curve to the observed
$J$-band flux-density. Due to the low continuum level in
this spectrum and relatively poor efficiency of GMOS at these
wavelengths, the constraint on the power-law continuum slope is fairly
poor with spectral indices in the range $0.5<\alpha_{\nu}<1.5$
acceptable. Planned infrared spectroscopy should yield a better value
for $\alpha_{\nu}$ and in this paper we adopt the `maximum continuum'
case where $\alpha_{\nu}=0.5$ as is typical for unreddened quasars.

With the continuum level defined as above, we measure the rest-frame
equivalent width of the observed \lya\ line to be 160 \AA. The main
uncertainty on this measurement is that of the continuum level. Since
we have adopted the maximum continuum level, this gives a lower limit
to the equivalent width. We also note that because of absorption of
the \lya\ line, the intrinsic equivalent width before absorption will
be even higher. A rest-frame equivalent width of $>160$ \AA\ for \lya\
is very high compared with that measured for a composite spectrum of
SDSS quasars of 93 \AA\ (Vanden Berk et al. 2001).   Another way of
comparing the \lya\ line to that of other quasars is the height of the
peak of the line compared to the continuum. For CFHQS\,J2329-0301, the
peak is 20 times the continuum level. This compares with a typical
peak height of only 3 times the continuum level in the SDSS composite.
All 19 quasars found at $z>5.7$ in the SDSS have peaks less than 10
times the continuum level (Fan et al. 2006b). Therefore it appears
that CFHQS\,J2329-0301 has an unusually strong narrow component to its
\lya\ line. The high \lya\ EW explains the unusual location of
this quasar in Fig.\,\ref{fig:izj} because \lya\ falls in the
$z'$ filter.

If the peak of the \lya\ line suffered from significant absorption,
then the intrinsic peak would be even higher. We consider this
unlikely given how high the \lya\ peak and equivalent width
are. Therefore we are confident that the measured \lya\ redshift of
$z=6.427 \pm 0.002$ is close to the systemic redshift of the quasar
and adopt $z=6.43 \pm 0.01$ as the redshift and its uncertainty. We
plan to obtain spectroscopy of the \mgii\ emission line to compare
with the \lya\ redshift, since \mgii\ usually shows only small offsets
from \lya\ (Richards et al. 2002). We note that previous work on
high-redshift quasars has usually defined the redshift via the \lya\
peak (Fan et al. 2006a and refs therein) and that the few high
redshift quasars with \lya\ and \mgii\ redshifts show no systematic
offset between them (Wyithe et al. 2005). 

\subsection{CFHQS\,J0033-0125}

This is the faintest of the four quasars with $z'=22.41$ which is
close to the CFHQS magnitude limit of $z'=22.5$. The spectrum
therefore has low S/N. In contrast to CFHQS\,J2329-0301,
CFHQS\,J0033-0125 has a very weak \lya\ emission line which barely
rises above the continuum. There is a narrow peak in \lya\ at
$\lambda=8674$\,\AA\ which we tentatively identify as the systemic
redshift giving $z=6.13$, but the relatively low height of this peak
above the continuum makes this redshift estimate quite
uncertain. Plausible values for the redshift lie in the range
$6.10<z<6.18$.

The equivalent width of the \lya\ emission is difficult to measure due
to the uncertain continuum but it appears to be $<10$\,\AA. With the
exception of this narrow peak, CFHQS\,J0033-0125 resembles the
lineless quasars discovered in the SDSS (Fan et al. 2006a and refs
therein). We suspect that the low equivalent width of \lya\ in this
quasar is due to an unusually high number of dense pockets of neutral
hydrogen close to the quasar.

Despite the low S/N, it is apparent that there is a higher transmitted
flux in the \lya\ forest at observed wavelengths 7500\,\AA\ --
8000\,\AA\ than at 8000\,\AA\ -- 8500\,\AA. This is observed in other
quasars at similar redshift (see for example CFHQS\,J1509-1749 on the
same figure) as expected based on the increase of the \lya\ forest
density with redshift and provides confidence that this object really
is a $z>6$ quasar and not an unusual broad absorption line quasar such
as have been found in the SDSS (Hall et al. 2002).

\subsection{CFHQS\,J1509-1749}

This quasar is the brightest discovered so far in the CFHQS with
$z'=20.26$. This magnitude is similar to the faintest $z \sim 6$ SDSS
quasars. The optical spectrum is the most sensitive of the four
presented here resulting in the highest S/N. The spectrum shows a strong continuum break across the broad
\lya\ emission line. This line has a peak at about 8660\,\AA\
($z=6.12$) but is highly asymmetric with considerably more flux on the
red side of this peak than the blue side due to neutral hydrogen
absorption.  The \lya\ emission line has a rest-frame equivalent width
of 48\,\AA\ which is typical of high-redshift quasars (unlike
CFHQS\,J0033-0125 and CFHQS\,J2329-0301). A more detailed discussion
of this optical spectrum is given in Sec.\,\ref{igm}.

\begin{figure}
\resizebox{0.48\textwidth}{!}{\includegraphics{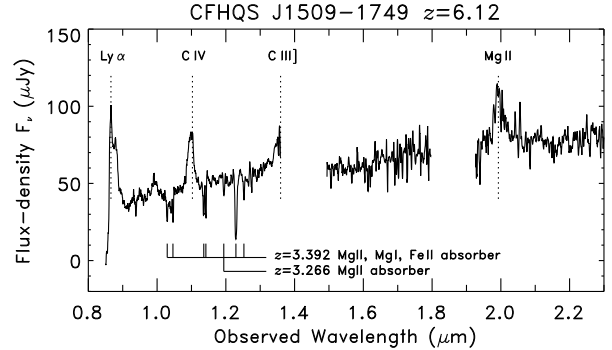}}
\caption{Near-IR spectrum of CFHQS\,J1509-1749 obtained with
  GNIRS. Several broad emission lines are labeled. Also shown are
  two metal absorbing systems at $z=3.266$ and 3.392. Note that this
  spectrum has been smoothed so the \mgii\ doublets, which are
  resolved in the original spectrum, appear as single lines. 
\label{fig:gnirs}
}
\end{figure}

The near-infrared spectrum of CFHQS\,J1509-1749 is shown in
Fig.\,\ref{fig:gnirs}. The gaps at $1.4\,\umu$m and $1.9\,\umu$m are
due to opaque wavelength regions of the Earth's atmosphere. Emission
lines of \lya, \civ, \ciii\ and \mgii\ are visible along with some
weaker lines. Also evident are strong absorption lines in the
$J$-band, which are discussed in more detail below.

The \mgii\ emission line is a low-ionization line that shows
significantly less asymmetry and blueshift (relative to the narrow
\oiii\ line) than high-ionization lines such as \civ\ (Richards et
al. 2002). Therefore it provides the best indicator of the systemic
redshift of a quasar at redshifts too high for any narrow lines to be
observable. We have performed a gaussian fit to the \mgii\ line
profile assuming an underlying power-law continuum. Although it is
known that the continuum around \mgii\ is not flat due to many blended
weak \feii\ lines, this does not affect the determination of the peak
wavelength of a fitted gaussian. The peak of the gaussian for \mgii\
is at $1.9923 \pm 0.0015\,\umu$m giving a redshift of $z=6.118 \pm
0.006$. We will hereafter refer to the redshift as $z=6.12$. Note that
this redshift agrees with the peak of \lya\ in both the GMOS and GNIRS
spectra. The \civ\ line is substantially offset from the \mgii\ and
\lya\ peaks. The measured \civ\ peak is at $1.0982 \pm 0.0010\,\umu$m
which is a blueshift from \mgii\ of $1200\,{\rm km\,s}^{-1}$.
Blueshifts of this size are common among lower redshift quasars
(Richards et al. 2002).

\subsubsection{Absorption lines}

Despite the moderate S/N ratio ($\sim 10$) of the near-infrared
spectrum of CFHQS\,J1509-1749, two strong absorption systems of low
ionization are clearly detected at $z=3.266 \pm 0.001$ and $3.392 \pm
0.001$. The spectral resolution is $R\simeq 900$, thus the two \mgii\
doublets are well resolved. They also are fully saturated as both
lines of each doublet have roughly the same equivalent width: $w_{\rm
obs}(2796,2803)=5.0$ and 21.0\,\AA\ for the $z=3.266$ and 3.392
systems, respectively. The $3\sigma$ equivalent width detection limit
in the $J$-band is $w_{\rm obs,lim}=4.0$\,\AA, thus even the weaker
\mgii\ system is well detected, at about the $4\sigma$ level.

The $z=3.392$ system is extremely strong, $w_{\rm
 rest}(2796)=4.8$\,\AA, with associated absorptions from the \mgi\
 singlet, $w_{\rm rest}(2852) =1.5$\,\AA, and \feii\ lines (2344,
 2382, 2586, 2600\,\AA), $w_{\rm rest}(2600)=2.9$\,\AA. This set of
 absorption lines is characteristic of damped \lya\ absorbers (DLAs):
 about 80\% of the \mgii\ systems with $w_{\rm rest}(2796) >1.0$\,\AA\
 are DLAs (Rao et al. 2006) and, moreover, all systems with $w_{\rm
 rest}(2852) >0.7$\,\AA\ are DLAs (Rao \& Turnshek 2000). This system
 has all the characteristics of a DLA.  The $z=3.266$ system has an
 \mgii\ doublet with $w_{\rm rest}(2796)=1.2$\,\AA, but no other
 absorption lines are detected.

The \mgii\ doublet is detectable in this spectrum over the redshift
range $2.10\leq z \leq 3.83$. This yields an incidence of $w_{\rm
rest}(2796)>1.0$\,\AA\ systems of d$N$/d$z\simeq 1.2$ for this single
line of sight. This value of d$N$/d$z$ is three times greater than the
results from \mgii\ absorption studies at $z\sim2$ which give
d$N$/d$z=0.40\pm 0.05$ for $w_{\rm rest}(2796)>1$\,\AA\ systems
(Nestor et al. 2005; Prochter et al. 2006). Given the
high Poisson noise, we cannot say that there is a statistically
significant excess in absorption towards CFHQS\,J1509-1749. We do not
expect significant cosmic evolution of d$N$/d$z$ for strong \mgii\
absorbers from $z=2$ to $z=3.4$, because only a weak evolution, if
any, is observed at $1.0<z<2.3$. Therefore the apparent excess towards
this quasar, if significant, must be due to a different cause, perhaps
related to the higher quasar redshift.

To determine if this discrepancy is statistically significant, one can
use the published results of Keck-ESI spectra of $z\sim6$ SDSS
quasars.  Strong \mgii\ doublets are detected towards SDSS\,J1044-0125
at $z=2.2786$ and SDSS\,J1306+0356 at $z=2.20$ and $2.53$, while there
is no \mgii\ absorption towards SDSS\,J0836+0054, SDSS\,J1030+0524 and
SDSS\,J1148+5251 (Becker et al. 2001; Djorgovski et al. 2001; Fan et
al.  2003; White et al. 2003; Ryan-Weber et al. 2006). The two \mgii\
systems of highest $z$ have clear, associated \feii\ (2586, 2600)
strong absorption, as could be seen from the quasar spectra, although
this is not mentioned by the authors. No equivalent width is reported
for any of the \mgii\ lines, but $w_{\rm obs}(2796)$ should be of
order 5~\AA\ or higher for the three systems.  The sum of the redshift
ranges for each quasar over which strong \mgii\ absorption is
detectable equals $\Delta z= 2.79$ which yields d$N$/d$z\simeq 1.08$.
A similar value is obtained when adding the two systems towards
CFHQS\,J1509$-$1749 to the sample, d$N$/d$z\simeq 1.11$ (or
$0.61<$\,d$N$/d$z<1.60$ at the $1\sigma$ level). Thus we see tentative
evidence for an enhancement in the absorber incidence for the most
distant quasars.

One possible explanation for this is that some of the most luminous $z
\sim 6$ quasars are gravitationally lensed (Wyithe \& Loeb 2002;
Comerford et al. 2002). The high incidence of \mgii\
absorption would then be related to above average mass density along
the line of sight. However, searches for multiple images and lensing
galaxies in the highest redshift SDSS quasars do not show any evidence for
lensing (Fan et al. 2003; Richards et al. 2004; Willott et al. 2005b).
There is no evidence for multiple images from the ground-based imaging
of the CFHQS quasars discovered so far (Fig.\,\ref{fig:cutouts}).


\subsubsection{Dust reddening}

\begin{figure}
\resizebox{0.48\textwidth}{!}{\includegraphics{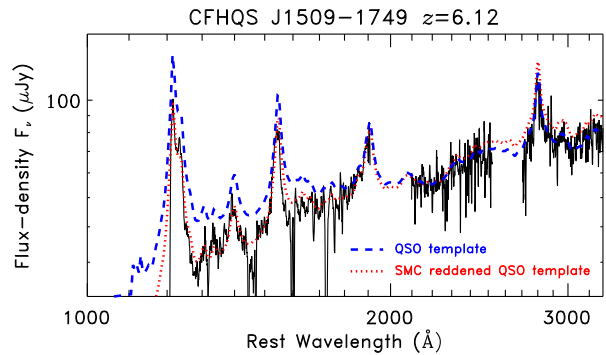}}
\caption{Rest-frame UV spectrum of CFHQS\,J1509-1749 plotted on a
  log-log scale. At $\lambda_{\rm rest}>2000$\,\AA\ the spectrum is
  well fit by an unreddened quasar template (blue dashed line). But at
  $\lambda_{\rm rest}<2000$\,\AA\ the spectrum is considerably
  reddened and can be fit by a quasar template reddened by SMC
  dust (red dotted line). Note that this SMC dust reddened quasar
  overpredicts the flux at $\lambda_{\rm rest}>2500$\,\AA.
\label{fig:redq}
}
\end{figure}

The full near-IR coverage of the GNIRS spectrum allows one to
investigate the continuum shape of this quasar. A power-law fit to the
$J$-band continuum gives a spectral index of $\alpha_{\nu}=1.00 \pm
0.02$. This is considerably redder than typical
quasars which have $\alpha_\nu=0.5$ (Richards et al. 2003).  However,
inspection of the $H$- and $K$-band spectra show that the spectrum
flattens here.

Because of the complex shape of quasar spectra at $\lambda_{\rm
rest}>2300$\,\AA\ due to blended \feii\ lines and Balmer continuum
(the ``small blue bump''), we do not fit a power-law to these data but
instead overplot a composite quasar spectrum constructed from spectra
of SDSS quasars (Fig.\,\ref{fig:redq}). This shows that a typical
quasar spectrum fits the continuum well in the $H$- and $K$-bands but
not at shorter wavelengths ($\lambda_{\rm rest}< 2000$\,\AA). An
analysis of extinction curves of reddened SDSS quasars (Hopkins et
al. 2004) has shown that quasar dust is most often similar to the dust
in the Small Magellanic Cloud (SMC). This extinction curve steepens
significantly in the UV compared to that of typical Milky Way (MW)
dust. We also plot on Fig.\,\ref{fig:redq} the composite quasar
spectrum reddened by enough SMC type dust ($A_V=0.10$) to match the
$J$-band spectrum. Although this reddened quasar template fits well at
$<2000$\,\AA, it predicts a steeper slope at $>2000$\,\AA\ than is
observed. We observe essentially no reddening of the intrinsic quasar
slope over the range $2000$\,\AA\ - 3200\,\AA. We have checked the
possibility of reddening by dust with known extinction laws (e.g. MW,
LMC, SMC) in the $z\sim3$ absorbers and found fits no better than SMC
dust at the quasar redshift.

One possible explanation for the sharp change in extinction at
$\lambda_{\rm rest} \approx 2000$\,\AA\ is that the dust reddening
this quasar has been created by type II supernovae. In our galaxy, the
dominant dust production mechanism is in low-mass AGB stars (Gerhz
1989). But at $z>6$ there is insufficient time ($<1$\,Gyr) for such
stars to have evolved off the main sequence. Therefore core-collapse
supernovae (whose progenitors have short main sequence lifetimes) may
be the main dust producers in the early universe. The heavy metal
content of supernovae explosions is expected to produce vast
quantities of dust with a quite different composition and grain size
distribution to that from AGB stars (Todini \& Ferrara 2001). Recent
observations of a $z>6$ broad absorption line quasar and a $z>6$ gamma ray burst have
both shown signs of extinction curves which are similar to that
inferred here (Maiolino et al. 2004; Stratta et al. 2007). However,
without rest-frame optical observations, it is impossible to say if
this is due to supernovae dust (which would have considerable, but
constant, extinction at $2000$\,\AA\ - 3000\,\AA; Maiolino et
al. 2004) or dust with only very small grains (so no extinction at all
at $>2000$\,\AA). 

We conclude that CFHQS\,J1509-1749 shows unusual dust reddening.
Infrared observations probing the rest-frame optical are necessary to
determine if this dust matches the predicted extinction curve of dust
created by type II supernovae.

\subsection{CFHQS\,J1641+3755}

The optical spectrum of this quasar (Fig.\,\ref{fig:spec}) shows a
broad \lya\ line with a sharp drop in flux on the blue side. The \lya\
line has typical asymmetry with the peak close to the sharp drop. We
assume the peak at $\lambda=8555$\,\AA\ defines the systemic redshift
of $z=6.04$. The redshift uncertainty is $\pm 0.04$. The spectrum of
CFHQS\,J1641+3755 is affected by night sky subtraction residuals
caused by incomplete removal of fringing which lead to excess positive
flux. This explains the apparent high transmission of the \lya\ forest
for this quasar.  Because of this we cannot with certainty define the
continuum level of this quasar redward of \lya. Given the fact that
the quasar has a $z'-J$ colour typical of quasars
(Fig.\,\ref{fig:izj}) we assume a spectral index of $\alpha_\nu=0.5$
to calculate the absolute magnitude from the $J$-band magnitude.

\section{Multi-wavelength data}
\label{multi}

We have searched the NASA/IPAC Extragalactic Database (NED) at the
locations of the four new quasars and found no coincident objects from
any other survey. Two of the quasars, CFHQS\,J0033-0125 and
CFHQS\,J1641+3755 are located in regions of the FIRST radio survey
(Becker et al. 1995). We have checked the FIRST image
cutouts and neither quasar is detected to a flux limit of $\approx
0.5$\,mJy meaning that neither quasar is radio-loud.

\subsection{1.2\,mm MAMBO observations. Dust and ultra-luminous IR starbursts.}

Follow-up observations of large samples of $2 \ltsimeq z \ltsimeq 6$
quasars at mm and sub-mm wavelengths have shown that a large fraction
(30\%) of optically luminous quasars are hyper-luminous infrared
($L_\mathrm{FIR} \gtsimeq 10^{13} \, L_\odot$) sources.  In these
sources, the far-IR luminosity is mainly related to the warm ($40-60
\, \mathrm{K}$) dust, with estimated dust masses of few $10^8 \, \rm
M_{\odot}$. The heating of the warm dust appears to be dominated by
the starburst activity of the quasar host galaxy and the implied star
formation rates are of $\rm \sim 1000 \, M_\odot \, yr^{-1}$ (e.g.
Omont et al. 2001; 2003).  In a growing number of cases, warm and
dense molecular gas is detected via CO emission lines and, in a few
cases, in other species, revealing the presence of large reservoirs of
molecular gas, the fuel of the star forming activity (see reviews by
Cox et al. 2005 and Solomon \& Vanden Bout 2005). The presence of
such huge starbursts in phases of strong accretion of the quasars
proves the simultaneity of major phases of growth of the most massive
(elliptical) galaxies and their super-massive black holes, and is an
important clue for explaining the black hole -- spheroid relation.

The detection of mm or sub-mm dust emission has been reported in eight
$z > 5.7$ SDSS quasars (but only three at $z \ge$\,6) with typical
flux densities at 1.2\,mm of a few mJy (Bertoldi et al. 2003a; Priddey
et al. 2003; Robson et al. 2004; Beelen et al. 2006; Wang et
al. 2007). Their properties and detection rate, as well as the CO
detection in J1148+5251 (Bertoldi et al. 2003b; Walter et al. 2003,
2004), are comparable to other high-$z$ strong starburst quasars. This
shows the presence of giant starbursts and vast amounts of molecular
gas, with heavy elements, less than one billion years after the Big
Bang. J1148+5251 is also the first high-$z$ source where \cii\ line
emission was detected (Maiolino et al. 2005).
	
It is important to extend the exploration of the millimeter properties
and star formation to a larger sample of $z \gtsimeq 6$ quasars,
especially among the more numerous, fainter quasars in the CFHQS. A
crucial question for building up the black hole -- spheroid relation is
whether the 1.2\,mm emission and hence the star formation rate are
strongly dependent on the UV luminosity, or only weakly as at $2<z<4$
(Beelen 2004; Cox et al.\ 2005; Wang et al.\ 2007). Therefore we have
observed at 1.2\,mm the four $z>6$ CFHQS quasars.

The observations were performed within the pool observing sessions at
the IRAM 30m telescope in Winter 2006/2007, using the 116 element
version of the Max Planck Millimeter Bolometer (MAMBO) array (Kreysa
et al. 1998) operating at a wavelength of 1.2\,mm (250\,GHz).  We used
the standard on-off photometry observing mode, chopping between the
target and sky at 2 Hz, and nodding the telescope every 10 or 20 s
(see e.g. Wang et al. 2007).  On average, the noise of the channel
used for point-source observations was about 35-40 $t^{-1/2}$\,
mJy\,beam$^{-1}$ (where $t$ is the integration time in seconds). This
allowed us to achieve rms $\ltsimeq 0.5$\,mJy for each of the four
sources, with about 1.5\,hr of telescope time per source. The data
were reduced with standard procedures to minimize the sky noise with
the MOPSIC package developed by R. Zylka.

\begin{table}                                     
\begin{center}
{\caption{\label{tab:mambo} MAMBO 1.2\,mm photometry of CFHQS quasars.}}
\begin{tabular}{lcc}  
\hline
\hline
\mc{1}{c}{Quasar} &\mc{1}{c}{Redshift} &\mc{1}{c}{$S_{1.2}$ (mJy)} \\
\tableline
CFHQS\,J0033-0125 & 6.13 & 1.13$\pm$0.36 \\
CFHQS\,J1509-1749 & 6.12 & 1.00$\pm$0.46 \\
CFHQS\,J1641+3755 & 6.04 & 0.08$\pm$0.46 \\
CFHQS\,J2329-0301 & 6.43 & 0.01$\pm$0.50 \\
\hline
\end{tabular}                
\end{center}                     
\end{table}

The results are shown in Table \ref{tab:mambo}. There is no strong
source with a flux significantly larger than 1\,mJy. CFHQS\,J0033-0125
is detected at S/N\,$>3$ with flux 1.13$\pm$0.36\,mJy. In addition,
CFHQS\,J1509-1749 has a very marginal (S/N\,$\sim$\,2) detection. The
mean flux in the four sources is 0.56$\pm$0.23\,mJy. This is an order
of magnitude weaker than J1148+5251. It corresponds nevertheless to a
far-infrared luminosity $\gtsimeq 10^{12} \,L_\odot$ and star
formation rate of $\rm \sim 300 \, M_\odot \, yr^{-1}$, typical of
ultraluminous infrared galaxies. This mean 1.2\,mm flux for our four
CFHQS sources is significantly smaller than the mean value that one
can deduce from Tables 1 and 2 of Wang et al.\ (2007) for 17
z\,$\sim$\,6 SDSS quasars which is $\approx$\,1.3\,mJy. This
difference could be related to the fainter UV luminosity of the CFHQS
quasars. Observations of more CFHQS quasars in the future will allow
tighter constraints on the luminosity-dependence of far-infrared
emission in the most distant quasars.

\section{Ionization state of the $z>5$ IGM}
\label{igm}

The spectra of the quasars presented here are discovery spectra. They
generally have modest integration times. Considering the faintness of
most of these quasars, detailed analysis of the implications of these
quasars for cosmic reionization is deferred until we have higher S/N
and resolution spectroscopy. In this section, we describe the
information we can get from our observations on the ionization state
of the high redshift universe. We only discuss the highest redshift
quasar, CFHQS\,J2329-0301, and the brightest quasar,
CFHQS\,J1509-1749, since the other two quasars have very low S/N
spectra. In addition, the systemic redshifts of CFHQS\,J0033-0125 and
CFHQS\,J1641+3755 are poorly determined due to the lack of near-IR
spectroscopy and low S/N of broad \lya. We note that poor sky
subtraction can be a serious problem for studies of the high-redshift
IGM. The spectra we use in this section are from GMOS in the
nod-and-shuffle mode which is well-documented to provide excellent sky
subtraction (Abraham et al 2004).

\subsection{Intrinsic quasar spectra}
\label{intrinsic}

In order to determine the neutral hydrogen absorption in these quasar spectra, we first need to determine the intrinsic spectra before absorption. This is not a trivial task given the large amount of absorption shortward of and around the \lya\ emission line. As described in Sec.\,\ref{indiv}, we used optical/near-IR spectroscopy and/or photometry to determine the continuum power-law indices at $\lambda_{\rm rest}>1250\,$\AA. CFHQS\,J1509-1749 has $\alpha_\nu=1.0$. The continuum for CFHQS\,J2329-0301 is not well-defined so we adopt $\alpha_\nu=0.5$. Observations of a large sample of quasars at lower redshift show a continuum break in the region  $1200\,$\AA\ $<\lambda<1300$\,\AA\ with a steeper power-law index of $\alpha_\nu=1.7$ at shorter wavelengths (Telfer et al. 2002). Therefore we adopt  $\alpha_\nu=1.7$ for both quasars at $\lambda_{\rm rest}<1250\,$\AA. 

Because we are interested in absorption close to the quasar we also include broad emission lines of \lyb, \lya\ and \nv. For CFHQS\,J1509-1749 we set the width of these lines in velocity to be equal to the measured \civ\ FWHM\,$=5600\,{\rm km\,s}^{-1}$ (consistent with the \mgii\ FWHM). The peak height of the \lya\ line is determined by eye and the \lyb\ and \nv\ lines have the same fraction of the \lya\ peak height as in the composite spectrum of Telfer et al. (2002). For CFHQS\,J2329-0301 the intrinsic width of the \lya\ emission line is very uncertain. The spectrum shown in Fig.\,\ref{fig:spec} has a strong narrow \lya\ line on top of a broad base. Whether this observed profile is due to absorption of a single broad line or a combination of broad+narrow lines is uncertain. We do not presently have near-IR spectroscopy for the measurement of other broad line widths. We adopt a single gaussian line with a width of FWHM\,$=3400\,{\rm km\,s}^{-1}$ and peak equal to the observed \lya\ peak. The emission lines only have a significant effect very close to the quasar redshift, so this uncertainty does not affect the results presented.

\begin{table}
\begin{center}
\caption{IGM absorption  in bins of width $\Delta z=0.15$ \\towards the $z=6.12$ quasar CFHQS\,J1509-1749 \label{tab:transtau}} 
\begin{tabular}{cccr}
\hline
\hline
\mc{1}{l}{Redshift} &\mc{1}{c}{Transition} &\mc{1}{c}{Transmission} &\mc{1}{c}{$\tau_{\rm Ly\alpha}$}\\
\tableline
5.225  &  \lya & $0.110 \pm 0.004$ & 2.20 \\
5.375  &  \lya & $0.084 \pm 0.005$ & 2.48 \\
5.525  &  \lya & $0.051 \pm 0.005$ & 2.97 \\
5.675  &  \lya & $0.026 \pm 0.005$ & 3.64 \\
5.825\tablenotemark{a}  &  \lya & $0.000 \pm 0.005$ & $>4.64$ \\
5.825\tablenotemark{b}  &  \lyb & $0.097 \pm 0.022$ & 5.25 \\
\hline
\end{tabular}
\tablenotetext{a}{Optical depth is a $2\sigma$ lower limit.}
\tablenotetext{b}{Transmission is after correction for foreground \lya\ as described in the text and the optical depth is the equivalent \lya\ optical depth using $\tau_{\rm Ly \alpha}=2.25\,\tau_{\rm Ly \beta}$.}  
\end{center}
\end{table}

\subsection{IGM optical depth}

The observed quasar spectra are divided by their intrinsic spectra as
defined in Sec.\,\ref{intrinsic} to determine the transmission as a
function of wavelength. The transmission is then used to calculate the
effective optical depth.  When binning over multiple pixels, each
pixel is weighted by the inverse variance to reduce the effects of
noise spikes due to bright sky emission lines. For CFHQS\,J2329-0301
it is evident that the low S/N of this discovery spectrum prevents us
from placing strong constraints on the optical depth of the IGM along
this line-of-sight. We detect flux in the region of \lya\ absorption
at $z<5.75$, but not at higher redshifts. In a large bin covering the
redshift range $5.35<z<5.75$ we measure a \lya\ optical depth of
$\tau=2.67 \pm 0.17$. In the higher redshift bin of $5.75<z<6.35$ we
obtain only a $2\sigma$ limit on the optical depth of
$\tau>3.93$. This lower limit for the optical depth is considerably
lower than the detections and limits for $z>6.2$ SDSS quasars, so we
do not discuss it further in this paper and await higher S/N
spectroscopy.

The spectrum of CFHQS\,J1509-1749 does have high enough S/N to
investigate the optical depth of the IGM. We bin the spectrum in
intervals of $\Delta z=0.15$ to get the effective optical depth along this
line-of-sight. We ensured that the highest redshift
bin does not include flux that may be associated with the quasar proximity zone discussed in
Sec.\,\ref{nearzone}. Values of \lya\ transmission and effective optical depth
are given in Table \ref{tab:transtau}. The uncertainties reflect only
the photon noise, which is the dominant cause of measurement
uncertainty. However, in most cases, this uncertainty is lower than the
intrinsic cosmic variance (Fan et al. 2006b). 

The \lyb\ transmission is measured only for the highest redshift
bin. Lower redshift bins of \lyb\ would overlap with the higher order
Lyman transitions. The \lyb\ region at $z=5.8$ is contaminated by
foreground \lya\ absorption at $z=4.7$. This foreground \lya\
absorption is corrected for statistically using the curve of effective
\lya\ optical depth against redshift measured for a large number of
lines-of-sight at $z<5.5$. Fan et al. (2006b) quote this as $\tau_{\rm
Ly \alpha}= 0.85[(1+z)/5]^{4.3}$. Applying this correction gives a
value for the \lyb\ transmission of $0.097\pm 0.022$. The quoted uncertainty
does not include the uncertainty in the foreground \lya\
transmission. For this bin we detect zero flux in \lya\ but there is a
$>4\sigma$ detection of flux in \lyb. This is because the oscillator
strength for the \lyb\ transition is considerably lower than for \lya.

\begin{figure}
\resizebox{0.48\textwidth}{!}{\includegraphics{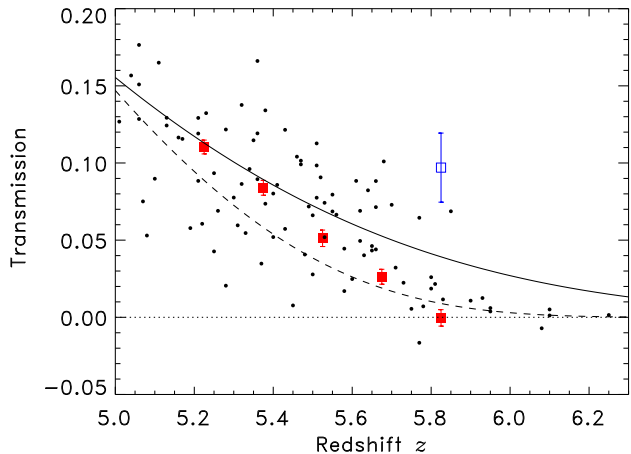}}
\vspace{0.2cm}
\caption{Lyman transmission in the spectrum of
 CFHQS\,J1509-1749. Solid (red) squares show the \lya\ transmission
 and the open (blue) square is the \lyb\ transmission in bins of width
 $\Delta z=0.15$. Small circles show the \lya\ transmission from the
 sample of 19 SDSS quasars of Fan et al. (2006b). The solid line is
 the fit to the \lya\ transmission of $z<5.5$ quasars given by Fan et
 al. (2006b). The dashed line was obtained by Becker et al. (2007) by
 fitting flux PDFs of $z<5.4$ quasars with a lognormal optical
 depth distribution.  At $z>5.7$ our data (and most of those in the SDSS) fall below the extrapolation of the Fan et al. curve.
\label{fig:trans}
}
\end{figure}

\begin{figure}
\hspace{-0.6cm}
\resizebox{0.50\textwidth}{!}{\includegraphics{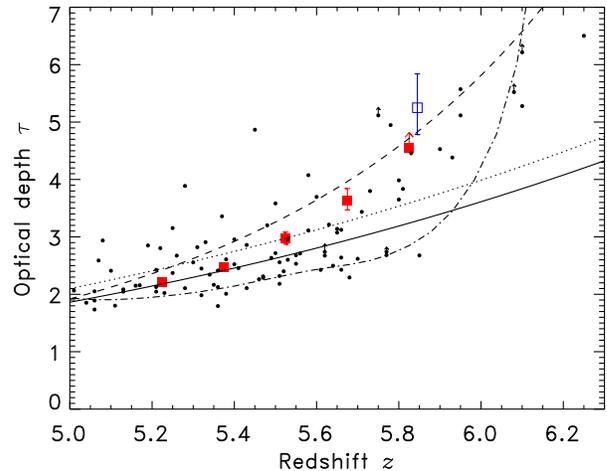}}
\vspace{0.2cm}
\caption{Effective Gunn-Peterson optical depth from the spectrum of
CFHQS\,J1509-1749 (symbols as in Fig.\,\ref{fig:trans}). The \lyb\
point (open blue square) is the equivalent \lya\ optical depth. The
solid and dashed lines are the extrapolations from lower redshift
quasars by Fan et al. (2006b) and Becker et al. (2007) (as in
Fig.\,\ref{fig:trans}). The dotted and dot-dashed lines are the
effective optical depth evolution in the simulations of Paschos \&
Norman (2005) and Gnedin \& Fan (2006), respectively.
\label{fig:tau}
}
\end{figure}

Fig.\,\ref{fig:trans} plots transmission against redshift for the five
\lya\ bins and one \lyb\ bin.  Also plotted are \lya\ transmissions
for SDSS quasars in Fan et al. (2006b). The solid curve is an
extrapolation of the fit to $z<5.5$ quasars described above. This
curve is consistent with the increasing absorption at higher redshifts
being solely due to the increase in number density of \lya\ forest
clouds for typical model IGM parameters (Miralda-Escud\'e et al. 2000;
see also the model curves in Fig.\,\ref{fig:tau})). The \lya\ data for
CFHQS\,J1509-1749 lie close to the curve at $z<5.4$ but fall well
below it at higher redshifts, indicating a sharp evolution in the
hydrogen neutral fraction compared to the expected extrapolation.

Becker et al. (2007) have cast doubt on this form for the expected
evolution in transmission. They showed that the \lya\ transmitted flux
probability distributions of quasars at $2<z<5$ are better fit by
lognormal optical depth distributions than the distributions from
theoretical models such as Miralda-Escud\'e et al. (2000). Because the
transmitted flux for the highest redshift quasars primarily depends
upon the low optical depth tail of the distribution in voids, using
the correct form for the optical depth distribution is important. The
dashed curve in Fig.\,\ref{fig:trans} shows the expected
transmission evolution based on fits to quasars at $z<5.4$ by Becker
et al. It is clear that this curve falls much more rapidly than that
of Fan et al. (2006b) and is consistent with the $z>5.8$ data for no
evolution in the hydrogen neutral fraction. However, the Becker et
al. curve does not do a good job of fitting the data at $5.4<z<5.7$,
where it predicts transmission considerably lower than most of the
SDSS data points. It is beyond the scope of this paper to investigate
the cause of this discrepancy, but we note that Becker et al. used
linear fits to evolution parameters, so it is perhaps not surprising
that this linear fit cannot accomodate the data at all redshifts.

Fig.\,\ref{fig:tau} shows the optical depth evolution. For the highest
redshift bin, the \lya\ point is a $2\sigma$ lower limit. For the
\lyb\ point, we plot the equivalent \lya\ optical depth assuming
$\tau_{\rm Ly \alpha}=2.25\,\tau_{\rm Ly \beta}$ (Fan et al. 2006b;
Gnedin \& Fan 2006). The \lya\ limit is consistent with the \lyb\
detection. The line-of-sight to CFHQS\,J1509-1749 shows one of the
highest optical depth measurements at $z \approx 5.8$. As for the
previous plot, it is clear that there is a sharp evolution away from
the Fan et al. (2006b) low redshift curve at $z>5.4$. Assuming
standard IGM parameters (e.g. Fan et al. 2001), the optical depth
evolution for CFHQS\,J1509-1749 is consistent with an increase in the
IGM hydrogen neutral fraction of a factor of three from $z=5$ to
$z=6$.

Also plotted on Fig.\,\ref{fig:tau} are two simulation results for the
Gunn-Peterson optical depth evolution. The simulations of Paschos \& Norman (2005; dotted curve)
show no evolution away from the low redshift extrapolation in this
redshift range because reionization is 99\% complete by $z=6.4$. Our
data and that of the SDSS clearly show that the optical depth is
greater than in this model at $z>5.7$.

The dot-dashed curve is the optical depth evolution for the L8N256
simulation of Gnedin \& Fan (2006). In these simulations the overlap
period of reionization occurs at $z=6.2$. The emissivity at $z=6\pm
0.1$ in this model is set to match the SDSS observations, so we should
expect agreement at this redshift. This simulation predicts a very
sharp increase in the optical depth at $z>6$. As is evident from
Fig.\,\ref{fig:tau}, the actual evolution for CFHQS\,J1509-1749 and
SDSS quasars is somewhat more gradual than in this model. This
suggests that the period of overlap is likely more extended than in
the simple model. As highlighted by Gnedin \& Fan (2006), the optical
depth at the end of overlap is dominated by Lyman limit systems and
simulations need to both resolve very small scales and cover very
large scales. Both simulations gives $\tau$ evolution during the epoch
in which reionization is complete comparable to the Fan et al. (2006b)
curve and much flatter than the Becker et al. (2007) curve. We defer a
more detailed comparison of the observed optical depth evolution with
models until we have high S/N spectra of more CFHQS quasars.

\subsection{Highly-ionized near-zones}
\label{nearzone}

One of the difficulties of probing the hydrogen neutral fraction via
quasar spectroscopy is that Lyman series absorption is extremely
efficient. Because the density of the IGM increases at high redshifts,
a neutral fraction (by volume) of only $2\times 10^{-4}$ at $z=6$
gives an optical depth at \lya\ of $\tau \sim 5$ (e.g. Fan et
al. 2006b).

Therefore it is useful to identify methods which are more sensitive to
high neutral fractions. One of these which has been extensively
studied in recent years is using the size of the region around the
quasar which it has ionized. This is variously referred to as an H{\sc
II} region, Stromgren sphere or highly-ionized near-zone. We will
adopt the term near-zone as this does not imply that the surroundings
are significantly neutral. 

For a quasar embedded in a uniform density IGM, the size of the
ionization front from the quasar is proportional to $f_{\rm HI}^{-1/3}$, $\dot N^{1/3}$ and $t_{\rm Q}^{1/3}$ where \fhi\
is the fraction of hydrogen in the IGM that is neutral, $\dot N$ is
the number of ionizing photons emitted by the quasar per second and
$t_{\rm Q}$ is the age of the quasar activity (Cen \& Haiman
2000; Wyithe \& Loeb 2004). Note that the latter two terms could be combined into the total
number of ionizing photons emitted so far by the quasar. Therefore a
measurement of the size of the ionization front (from spectroscopy)
combined with knowledge of the number of ionizing photons emitted
would give the neutral fraction.

This method has been employed to argue that the highest redshift SDSS
quasars are embedded in a highly neutral IGM (\fhi$ \gtsimeq 0.1$;
Wyithe \& Loeb 2004; Wyithe et al. 2005; Mesinger \& Haiman
2007). However there are many complicating factors such as the
uncertain density distribution close to luminous quasars and
pre-ionization by associated galaxies (Yu \& Lu 2005). Uncertainty in
the systemic redshifts can also be an issue leading to a typical
uncertainty in the near-zone size of $\approx 1$\,Mpc for redshifts
from the \lya\ line.

An additional uncertainty comes from relating observations to
simulations.  Defining the exact end of a near-zone from spectra is
difficult. This is especially true for $5.7<z<6.2$ quasars which
typically do not show Gunn-Peterson troughs immediately shortward of
their near zones. In order to be able to define near-zone sizes for a
wide range of spectra, Fan et al. (2006b) adopted the definition as
the point in the spectra where the \lya\ transmission first dropped to
$T<0.1$ for spectra binned in 20\AA\ pixels. Most studies involving
just those quasars at the upper end of the SDSS redshift distribution
(e.g. Wyithe et al. 2005; Yu \& Lu 2005) define the near-zone size as
corresponding to the onset of the Gunn-Peterson trough (typically $T
\ltsimeq 0.01$). Bolton \& Haehnelt (2007a) discuss many of the
uncertainties associated with using near-zone sizes to estimate
\fhi. Chief among these is the fact that the observed size may not
correspond to the actual size of the ionized region due to significant
residual neutral hydrogen left behind the ionization front.

As mentioned previously, the sizes of near-zones are expected to be
related to the quasar luminosity. For the classical case of an
expanding \hii\ region the near-zone size is proportional to  $\dot N^{1/3}$. Bolton \& Haehnelt
(2007a) found the expected dependence to be $\dot N^{1/2}$ for
near-zones defined as $T<0.1$. The CFHQS magnitude limit is 2.3
magnitudes fainter than that of the SDSS meaning that the CFHQS quasars
will be on average almost a factor of 10 lower in
luminosity. Therefore the CFHQS near-zone sizes are expected to be
between $2 -3$ times smaller than those in the SDSS. Comparing the
near-zones sizes for the final CFHQS sample to the final SDSS sample
will allow a test of the luminosity dependence.

\begin{figure}
\resizebox{0.48\textwidth}{!}{\includegraphics{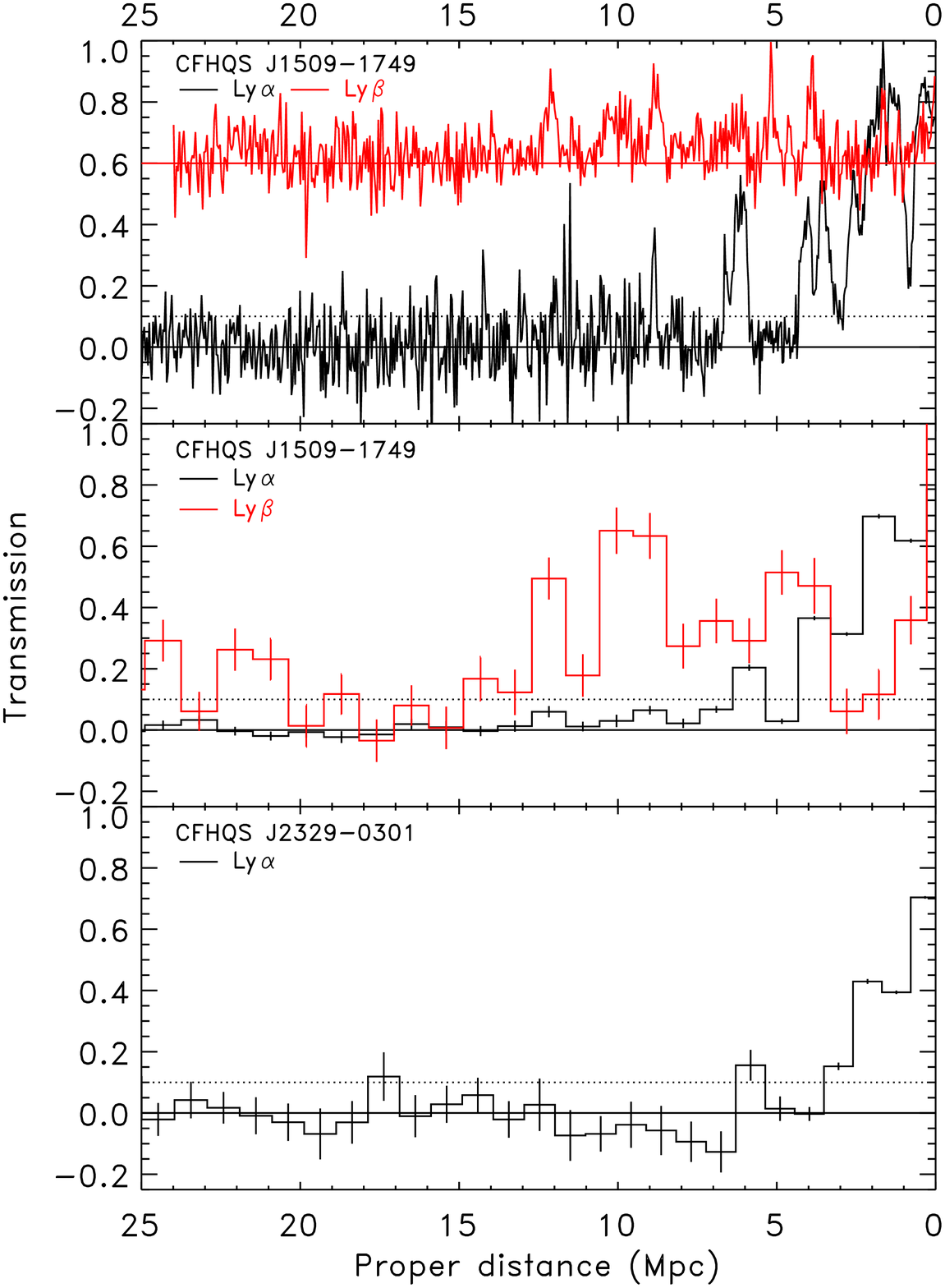}}
\caption{Lyman transmission close 
to the quasar redshift for CFHQS\,J1509-1749 (upper two panels) and
CFHQS\,J2329-0301 (lower panel). The horizontal scale shows the
distance in physical coordinates from the quasar. The upper panel
shows the \lya\ and \lyb\ unbinned spectra for CFHQS\,J1509-1749. The
\lyb\ transmission has not been corrected for foreground \lya\ absorption 
and is shown vertically offset by +0.6 for clarity. The middle panel
shows the same data but binned in 20\AA\ bins with \lyb\ now corrected
for foreground \lya\ absorption. The lower panel shows  20\AA\ binned
\lya\ transmission for CFHQS\,J2329-0301. Dotted horizontal lines show
$T=0.1$. 
\label{fig:nearzone}
}
\end{figure}

In the upper and middle panels of Fig.\,\ref{fig:nearzone} we show the
transmission in the spectrum of CFHQS\,J1509-1749 close to the quasar
redshift. The middle panel shows the \lya\ and \lyb\ transmission
binned in 20\,\AA\ bins. It is seen that the \lya\ transmission first
drops below $T=0.1$ at a distance of 4.3\,Mpc, but then rises up to
$T=0.2$ in the next bin and then falls once more below $T=0.1$ at
6.4\,Mpc. This behaviour of $T>0.1$ peaks beyond the distance where
$T$ first drops below 0.1 is also seen in the near-zones of some SDSS
quasars (Fan et al. 2006b) and in the simulated quasar line-of-sight
of Bolton \& Haehnelt (2007a) and is likely due to to the existence of
low-density regions within the quasar near-zone. Statistically
significant transmission below the $T=0.1$ level is detected in bins
at distances up to 12.7\,Mpc. Given these observations, what should we
conclude about the size of the highly-ionized near-zone surrounding
this quasar?

To further understand the properties of the near-zone we also show the
binned \lyb\ transmission in the middle panel of
Fig.\,\ref{fig:nearzone}. These data have been statistically corrected
for foreground \lya\ absorption. Excluding the first few Mpc, there is
statistically significant \lyb\ transmission all the way out to
12.7\,Mpc. In fact, the \lyb\ optical depth in this region is
considerably lower than expected given the observed \lya\ optical
depth, which suggests the foreground \lya\ forest at $z\approx
4.9$ along this line-of-sight is unusually transparent. 

The upper panel shows the \lya\ and \lyb\ transmission (this time
without statistical correction of the foreground \lya\ forest) of the
full resolution spectrum without binning. There is a correspondence
between positive features in \lya\ and \lyb\ at most places, although
the correspondence in width and height of these features vary due to the varying foreground \lya\ absorption. The transmission
feature at 12\,Mpc is clearly visible in \lyb. From the weakness of
this feature at \lya\ we conclude that the \lyb\ feature is due both
to a low-density region near the quasar and an unusually transparent
patch at $z\approx 4.9$. We note, as emphasized by Songaila (2004),
that if flux is detected at all at \lya, it is better to adopt this
measurement than to trust the apparently higher S/N \lyb\ measurement
because of uncertainty in the foreground subtraction. Nevertheless,
the detection of positive flux in all \lya\ bins out to 12.7\,Mpc
suggests that this is the true size of the zone impacted by the quasar
ionizing flux. This is significantly greater than the
6.4\,Mpc that would result from imposing a $T<0.1$ criterion.

Throughout the above discussion we have assumed that the flux detected
is not due to transparency caused by a very low neutral fraction
($f_{\rm HI}<10^{-4}$) in the IGM beyond the quasar ionization
front. There are a couple of facts supporting this assumption. First,
there is significant flux detected in every \lyb\ bin from 3 to
12.7\,Mpc and positive (though not always significant) \lya\ flux in
every bin. Then at larger distances, the IGM appears very opaque in
\lya\ and \lyb\ until there is some \lyb\ transmission at $\sim
20$\,Mpc. The transmitted flux in the region 6.4 to 12.7\,Mpc
corresponds to an effective \lya\ optical depth of $\tau=3.0$. This is
to be compared with the optical depth at similar redshift (this region
is $5.92<z<6.02$) and bin size along the lines-of-sight to all four
highest redshift SDSS quasars of $\tau>4.3$ (Fan et al. 2006b). If the
transmission at 6.4 to 12.7\,Mpc is beyond the quasar ionization
front, then the IGM here has an unusually high ionizing background
over a substantial redshift range.

The lower panel of Fig.\,\ref{fig:nearzone} shows the \lya\
transmission close to CFHQS\,J2329-0301. The spectrum does not have a
high enough S/N to provide useful \lyb\ data. The \lya\ near-zone is
remarkably similar to that of CFHQS\,J1509-1749 with the transmission
first dropping below $T=0.1$ at a distance of 3.7\,Mpc followed by a
spike with $T>0.1$ and then falling below $T=0.1$ again at
6.3\,Mpc. Due to the lower S/N of the data we would not be able to
detect any peaks at larger distances of comparable strength to those
in CFHQS\,J1509-1749.

The CFHQS quasars are mostly of lower luminosity than those of the
SDSS so we must scale our near-zone sizes to a common luminosity
before comparison with SDSS quasars and simulations carried out to
match SDSS quasars. We scale the sizes to the equivalent for
$M_{1450}=-27$ assuming $R\propto \dot N^{1/3}$.  Because
CFHQS\,J1509-1749 has $M_{1450}=-26.98$, no luminosity-scaling is necessary. For
CFHQS\,J2329-0301 with $M_{1450}=-25.23$, the near-zone size of
6.3\,Mpc goes up to 10.8\,Mpc after luminosity-scaling.

We can compare the luminosity-scaled sizes of the CFHQS near-zones with those
from the more luminous SDSS quasars. Bolton \& Haehnelt (2007b) give
the luminosity-scaled sizes of the \lya\ near-zones measured from the four SDSS
quasars at $z>6.1$ (excluding the broad absorption line quasar
SDSS\,J1048+4607) using a similar definition to us of the last pixel
where $T>0.1$. The four SDSS near-zone sizes are 2.6, 4.8, 6.2 and
13.6\,Mpc. Our two CFHQS sizes of 6.4 and 10.8\,Mpc are within the
range of the SDSS quasars, somewhat towards the upper end of the
distribution.

Although there are many uncertainties associated with using the
near-zone sizes to estimate the neutral fraction of the IGM that the
quasar ionization front expands into, the finding of relatively large
near-zone sizes in the two CFHQS quasars does allow us to determine
some constraints. Firstly, as shown by Bolton \& Haehnelt (2007a),
there is little dependence upon the quasar lifetime if $f_{\rm HI}
\ltsimeq 0.1$ and/or at $t_{\rm Q}\gtsimeq 10^7$\,yrs because in these
cases the observed near-zone size depends not on the expanding
ionization front, but on the residual neutral hydrogen left behind the
ionization front. Both short lifetimes and high neutral fractions lead
to small near-zone sizes, so we can safely assume that quasar lifetime
is not a factor. Bolton \& Haehnelt (2007b) simulated many quasar
lines-of-sight with parameters of $M_{1450}=-27$ and $t_{\rm Q} =
10^7$\,yrs. They found that for $f_{\rm HI} >0.3$, the near-zone sizes are all
$<4.5$\,Mpc. At $f_{\rm HI} = 0.1$, the average near-zone size is
$4.5$\,Mpc with a range up to $6.5$\,Mpc. The only simulated near-zone
sizes greater than $10$\,Mpc come from simulations with $f_{\rm HI}
<10^{-3}$.

The luminosity-scaled sizes for the two CFHQS quasar near-zones are
6.4 and 10.8\,Mpc.  Therefore we can be quite confident that neither
quasar is located in an IGM with $f_{\rm HI} >0.3$. For
CFHQS\,J2329-0301 at $z=6.43$ with a luminosity-scaled near-zone size
of 10.8\,Mpc, a very highly ionized IGM is inferred. Even if the
quasar redshift has been overestimated by $\delta z=0.02$, then the
observed near-zone size of 5.2\,Mpc would still give a
luminosity-scaled size of 9.0\,Mpc.  For both quasars, the quasar
radiation is emitted into a substantially pre-ionized IGM. This
indicates that these locations had already been reionized to $f_{\rm
HI} <0.3$ at their respective redshifts of $z=6.12$ and
$z=6.43$. Several authors have suggested that because luminous quasars
are expected to form in rare overdense regions, the surrounding IGM
had already been pre-ionized by galaxies (Yu \& Lu 2005; Alvarez \&
Abel 2007; Lidz et al. 2007). The possibility that quasar near-zones
do not represent typical regions of the IGM must always be kept in
mind when interpreting results such as these. However, we note that
the lower luminosities of most CFHQS quasars, such as
CFHQS\,J2329-0301, mean they are expected to be powered by less
massive black holes hosted in less rare, lower mass dark matter halos,
reducing the likelihood of pre-ionization by cluster galaxies.

\section{Conclusions}

We have presented the discovery of four new CFHQS quasars at
$z>6$. These were discovered from incomplete follow-up of less than
half the final survey area and we expect many more quasars to be found
in the next few years. We have discussed the observed properties of
the quasars including emission lines, intervening absorption lines,
continuum slopes and dust reddening. 

Sensitive millimeter continuum observations have been carried out with
MAMBO at the IRAM 30m telescope. One of the four quasars has been
securely detected. This strong dust continuum is likely due to a high
star formation rate in its host galaxy. The mean 1.2\,mm flux of the
CFHQS quasars is substantially lower than the mean 1.2\,mm flux of
SDSS quasars at similar redshifts. This is indicative of a
luminosity-dependence of the dust continuum emission.

We have investigated the constraints that the discovery spectra can
put on the ionization state of the IGM. For all but one quasar, higher
S/N spectroscopy are required to perform a full analysis of the IGM
properties.

CFHQS\,J1509-1749 at $z=6.12$ shows significant IGM optical depth
evolution from $z=5$ to $z=6$. The highest redshift bin at
$5.75<z<5.90$ has an effective optical depth of $\tau_{\rm Ly \alpha}\approx 5$,
which is one of the highest effective optical depths found at this redshift and
consistent with a factor of three increase in \fhi\ from $z=5$.

Both CFHQS\,J1509-1749 and CFHQS\,J2329-0301 have large
highly-ionized near-zones of at least 6 proper Mpc. These large near-zones
are inconsistent with quasar ionization fronts expanding into a highly
neutral ($f_{\rm HI} \gtsimeq 0.3$) IGM (Bolton \& Haehnelt 2007b). Our
results are broadly consistent with the (mass-averaged) value of $f_{\rm HI}
\sim 0.04$ at $z\sim6.4$ determined by Fan et al. (2006b) from
evolution of SDSS quasar near-zone sizes. Our strongest constraint
is from CFHQS\,J2329-0301 at $z=6.43$ which has a luminosity-scaled
near-zone size of 10.8\,Mpc which is only consistent with $f_{\rm HI}
< 10^{-3}$. In addition, the lower luminosity of this quasar compared
to SDSS quasars makes it less likely that its surroundings were
pre-ionized by cluster galaxies.

Taking together the results of (i) a significant increase in the IGM
optical depth at $z\approx 5.8$ (for CFHQS\,J1509-1749 and also
observed in the SDSS) and (ii) the low neutral fractions inferred from the
near-zones of CFHQS (and some SDSS)  $z>6.1$ quasars, suggest that reionization
is well under way before $z=6.4$ but is still not complete by
$z=5.7$. This is a more extended period than results from most simple
models where overlap occurs rapidly.

Finally we note that cosmic variance is still a significant issue for
the highest redshift quasars and that many more quasars at $z>6.1$ are
required to accurately determine the IGM evolution at $z>6$.

\acknowledgments

Based on observations obtained with MegaPrime/MegaCam, a joint project
of CFHT and CEA/DAPNIA, at the Canada-France-Hawaii Telescope (CFHT)
which is operated by the National Research Council (NRC) of Canada,
the Institut National des Sciences de l'Univers of the Centre National
de la Recherche Scientifique (CNRS) of France, and the University of
Hawaii. This work is based in part on data products produced at
TERAPIX and the Canadian Astronomy Data Centre as part of the
Canada-France-Hawaii Telescope Legacy Survey, a collaborative project
of NRC and CNRS. Based on observations obtained at the Gemini
Observatory, which is operated by the Association of Universities for
Research in Astronomy, Inc., under a cooperative agreement with the
NSF on behalf of the Gemini partnership: the National Science
Foundation (United States), the Particle Physics and Astronomy
Research Council (United Kingdom), the National Research Council
(Canada), CONICYT (Chile), the Australian Research Council
(Australia), CNPq (Brazil) and CONICET (Argentina). This paper uses
data from Gemini programs GS-2006A-Q-16, GS-2006A-Q-38 and
GS-2006B-Q-17. The Hobby-Eberly Telescope (HET) is a joint project of
the University of Texas at Austin, the Pennsylvania State University,
Stanford University, Ludwig-Maximillians-Universitaet Muenchen, and
Georg-August-Universitaet Goettingen.  The HET is named in honor of
its principal benefactors, William P. Hobby and Robert E. Eberly. The
Marcario Low Resolution Spectrograph was developed by the HET
partnership and the Instituto de Astronomia, Universidad Nacional
Autonoma de Mexico (IAUNAM). The LRS is named for Mike Marcario of
High Lonesome Optics who fabricated several optics for the instrument
but died before its completion. Based on observations made with the
ESO New Technology Telescope at the La Silla Observatory. Based on
observations with the Kitt Peak National Observatory, National Optical
Astronomy Observatory, which is operated by the Association of
Universities for Research in Astronomy, Inc. (AURA) under cooperative
agreement with the National Science Foundation. Based on observations
made with the Nordic Optical Telescope, operated on the island of La
Palma jointly by Denmark, Finland, Iceland, Norway, and Sweden, in the
Spanish Observatorio del Roque de los Muchachos of the Instituto de
Astrofisica de Canarias. Based on observations with the IRAM 30m MRT
at Pico Veleta. IRAM is supported by INSU/CNRS (France), MPG (Germany)
and IGN (Spain). This research has made use of the NASA/IPAC
Extragalactic Database (NED) which is operated by the Jet Propulsion
Laboratory, California Institute of Technology, under contract with
the National Aeronautics and Space Administration. Support for
M.G. was under REU program grant NSF AST-0243745. Thanks to the queue
observers at CFHT, Gemini, HET and IRAM who obtained data for this
paper. Thanks to Matthew Shetrone for coordinating the HET
observations. Thanks to JJ Kavelaars for advice on planning our MegaCam
observations in the CFHTLS Very Wide survey region. Thanks to Nick
Gnedin for providing the results of his simulations in electronic
form. Thanks to the anonymous referee for suggestions which improved
this paper.



\appendix

\section{Finding charts}

\begin{figure*}[b]
\hspace{-0.6cm}
\resizebox{0.99\textwidth}{!}{\includegraphics{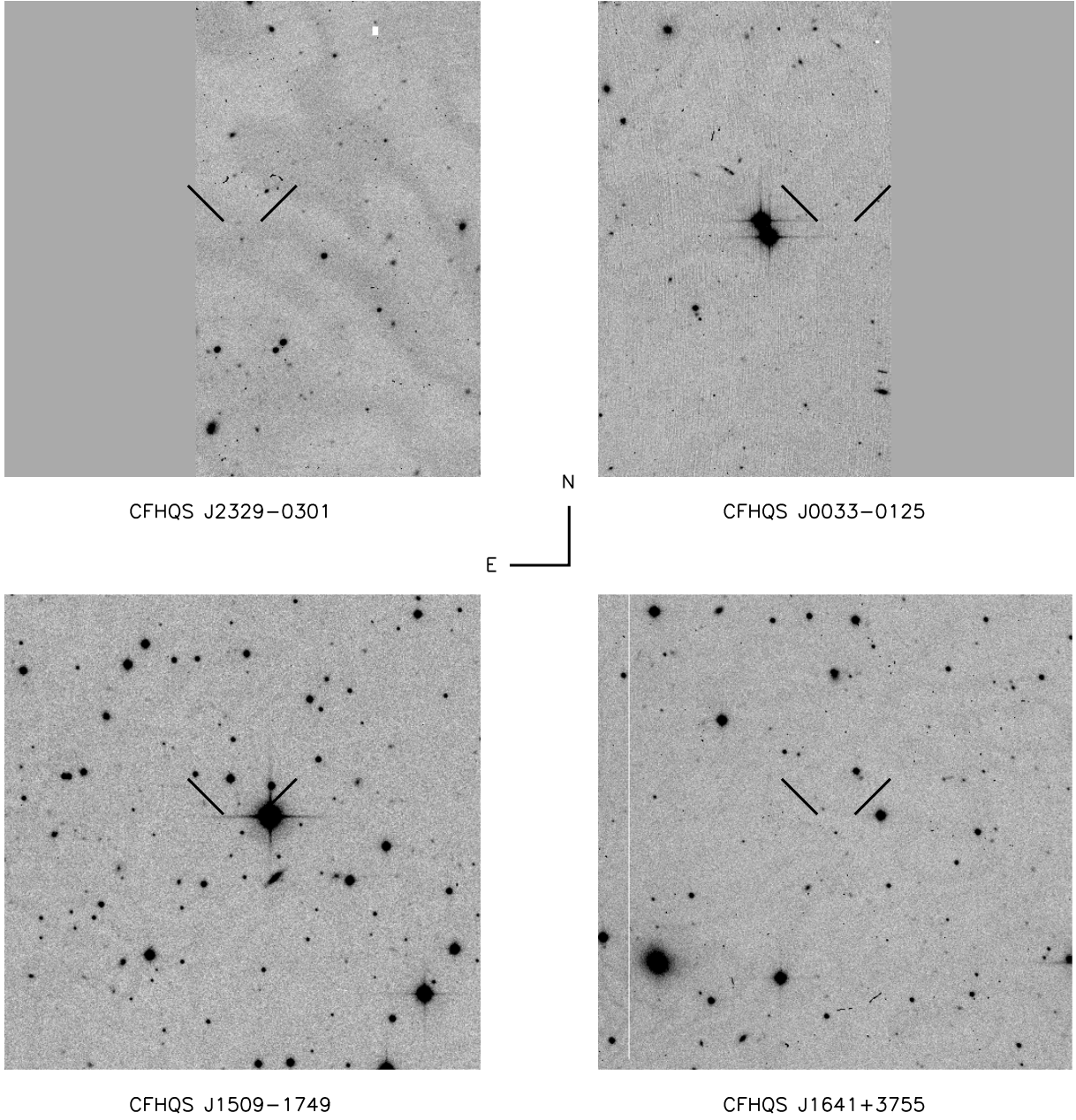}}
\caption{ $z'$-band finding charts for the CFHQS quasars.
\label{fig:finders}
}
\end{figure*}

Fig.\,\ref{fig:finders} presents $3' \times 3'$ finding charts for the
CFHQS quasars. All images are centred on the quasars and have the same
orientation on the sky. These are the MegaCam $z'$-band images in
which the quasars were first identified. MegaCam has gaps between the
CCDs and these data were not dithered so the gaps remain and are
evident in the upper two images where the quasars lies clos to the
edge of a CCD. For all quasars except CFHQS\,J1509-1749, the Megacam
data are a single exposure leading to many cosmic rays in the final images.

\end{document}